\documentclass[12pt,a4paper]{article}
\usepackage{epsfig}
\newcommand{\BR}{{\rm BR}}
\newcommand{\be}{\begin{equation}}
\newcommand{\ee}{\end{equation}}
\newcommand{\bea}{\begin{eqnarray}}
\newcommand{\eea}{\end{eqnarray}}

\begin{document}
\begin{titlepage}
\begin{center}
{\hbox to\hsize{\hfill }}

\vspace{4\baselineskip}

\textbf{\Large 
Recent MEG Results and Predictive SO(10) Models
}  
\bigskip
\bigskip
\vspace{2\baselineskip}

{\large
Takeshi Fukuyama$^{a,}$%
\footnote{E-Mail: fukuyama@se.ritsumei.ac.jp}
and 
Nobuchika Okada$^{b,}$%
\footnote{E-Mail: okadan@ua.edu}
} \\ 
\bigskip
\textit{\small 
$^a$Department of Physics, Ritsumeikan University, Kusatsu, 
Shiga 525-8577, Japan 
}
\bigskip
\textit{\small
$^b$Department of Physics and Astronomy, 
University of Alabama, Tuscaloosa, AL 35487, USA}

\vspace{3\baselineskip}

\textbf{Abstract}\\
\end{center}
\noindent
Recent MEG results of a search for the lepton flavor violating (LFV) 
 muon decay, $\mu \to e \gamma$, show 3 events 
 as the best value for the number of signals  
 in the maximally likelihood fit. 
Although this result is still far from the evidence/discovery 
 in statistical point of view, it might be a sign 
 of a certain new physics beyond the Standard Model. 
As has been well-known, supersymmetric (SUSY) models can generate 
 the $\mu \to e \gamma$ decay rate within the search reach 
 of the MEG experiment. 
A certain class  of SUSY grand unified theory (GUT) models 
 such as the minimal SUSY SO(10) model 
 (we call this class of models ``predictive SO(10) models'') 
 can unambiguously determine fermion Yukawa coupling matrices, 
 in particular, the neutrino Dirac Yukawa matrix. 
Based on the universal boundary conditions for soft SUSY breaking 
 parameters at the GUT scale, 
 we calculate the rate of the $\mu \to e \gamma$ process 
 by using the completely determined Dirac Yukawa matrix 
 in two examples of predictive SO(10) models. 
If we interpret the 3 events in MEG experiment 
 as a positive signal and combine it with other experimental constraints 
 such as the relic density of the neutralino dark matter 
 and recent results on muon $g-2$, we can pin down 
 a parameter set of the universal boundary conditions. 
Then, we propose benchmark sparticle mass spectra 
 for each predictive SO(10) model, which will be tested 
 at the Large Hadronic Collider. 

\end{titlepage}

\setcounter{footnote}{0}
\newpage
%
%

The search for new physics beyond the Standard Model (SM) 
 has been performed in a variety of energy scales. 
In addition to the direct searches for new physics 
 at the Large Hadronic Collider (LHC), being the collider experiment 
 with the highest energy at present, a search for the LFV processes 
 at low energies is also  very important in discovering new physics. 
This is because the LFV processes are highly suppressed in the SM, 
 and any positive signal once observed can be an evidence of  
 new physics.

Recently MEG collaboration \cite{MEG} has reported new results of 
 a search for the $\mu \to e \gamma$ decay and a maximally 
 likelihood analysis sets an upper limit at 90\% C.L. on 
 the branching ratio, $\BR (\mu \to e \gamma) < 1.5 \times 10^{-11}$, 
 which is at the same level of the current smallest limit 
 set by the MEGA experiment \cite{MEGA}, 
 $\BR (\mu \to e \gamma) < 1.2 \times 10^{-11}$.  
Very interestingly, the results  of the MEG experiment also show 
 3 events as the best value for the number of signals 
 in the maximally likelihood fit, which corresponds to \cite{MEG}
\bea 
 \BR (\mu \to e \gamma) = 3 \times 10^{-12}   
 \label{MEGsignal}
\eea
 for the center value. 
Although this result is still far from the evidence/discovery 
 in statistical point of view, it might be a sign of 
 a certain new physics beyond the Standard Model.

Non-zero masses and flavor mixings of neutrinos  
 observed through the neutrino oscillation phenomena
 imply that the lepton flavor in each generation is not 
 individually conserved, and therefore, 
 LFV processes among the charged-leptons exist. 
However, in simply extended models of the SM 
 so as to incorporate massive neutrinos,  
 the rate of the LFV processes are strongly suppressed 
 and far out of the reach of the experimental detection. 
This is because of the GIM mechanism which leads to 
 a suppression factor of the ratio between the neutrino mass 
 scale and the electroweak scale. 
It has been known \cite{LFVSUSY1, LFVSUSY2} 
 that SUSY models can generate the experimentally detectable rate 
 of the LFV processes through the LFV sources 
 in soft SUSY breaking terms.

The minimal supersymmetric Standard Model (MSSM) is one 
 of the most promising candidate for new physics, 
 providing a solution to the gauge hierarchy problem 
 of the SM. 
In addition, the successful gauge coupling unification at 
 $M_{GUT} \simeq 2 \times 10^{16}$ GeV supported by 
 the low energy data of the SM gauge couplings 
 strongly suggests the paradigm of SUSY GUT. 
Thus, the SUSY GUT models with unified gauge groups are 
 well-motivated theories at high energies. 
Among them, models based on the gauge group SO(10) are 
 probably the most compelling ones in terms of neutrino physics, 
 because quarks and leptons in each generation are unified 
 into a single ${\bf 16}$ representation along with 
 a right-handed neutrino, and the smallness of the neutrino masses 
 can be naturally explained through the seesaw mechanism \cite{seesaw}. 
Furthermore, a class of SO(10) models such as 
 the minimal SUSY SO(10) model \cite{BM} 
  has a very interesting feature.  
Because of the complete unification of quarks and leptons 
 into a single ${\bf 16}$ representation and introduction 
 of the minimal set of Higgs multiplets, fermion Yukawa matrices 
 are highly constrained, and it is very non-trivial to fit all the current  
 data of fermion masses and mixing angles including the neutrino sector. 
It has been shown \cite{Fukuyama-Okada} that the minimal SO(10) model can 
 simultaneously reproduce all the observed quark-lepton mass matrix
 data involving the neutrino oscillation data. 
It is very interesting that after the data fitting, 
 no free parameters are left and hence, 
 all the fermion Yukawa matrices, in particular, 
 the neutrino Dirac Yukawa matrix, are unambiguously determined.

In this paper, we consider such a class of SUSY SO(10) models 
 (predictive SO(10) models). 
We assume a suitable Higgs sector which breaks SO(10) 
 to the MSSM gauge group at the GUT scale 
 with generating masses for the right-handed neutrinos. 
Below the GUT scale, the low energy effective theory of the model 
 is described as the MSSM with the right-handed neutrino chiral
 multiplets. 
In this effective theory, the superpotential 
 in the leptonic sector is given by 
\begin{eqnarray}
W_Y =  Y_{\nu}^{ij} (\nu_R^c)_i \ell_j H_u 
+ Y_e^{ij} (e_R^c)_i \ell_j H_d 
+ \frac{1}{2} M_{R_{ij}} (\nu_R^c)_i  (\nu_R^c)_j 
   + \mu H_d  H_u  , 
 \label{superpotential}
\end{eqnarray} 
where the indices $i$, $j$ run over three generations, 
 $H_u$ and $H_d$ denote the up-type and down-type MSSM Higgs doublets, 
 respectively, and  $M_{R_{ij}}$ is 
 the heavy right-handed Majorana neutrino mass matrix. 
We work in the basis where the charged-lepton Yukawa matrix 
 $Y_e$ and the mass matrix $M_{R_{ij}}$ 
 are real-positive and diagonal matrices:  
 $Y_e^{ij}=Y_{e_i} \delta_{ij}$ and  
 $M_{R_{ij}}=\mbox{diag} ( M_{R_1},  M_{R_2},  M_{R_3}) $. 
Thus the off-diagonal components of the neutrino Dirac Yukawa 
 coupling matrix $ Y_{\nu}$ break the conservation of 
 the lepton flavor in each generation. 
Note that once we fix a predictive SO(10) model, 
 all components in the neutrino Dirac Yukawa matrix $ Y_{\nu}$ 
 and the mass spectrum of the right-handed neutrinos are 
 completely determined. 
The soft SUSY breaking terms in the leptonic sector 
 is described as 
\begin{eqnarray}
 -{\cal L}_{\mbox{soft}} &=& 
   \tilde{\ell}^{\dagger}_i 
   \left( m^2_{\tilde{\ell}} \right)_{ij}
   \tilde{\ell}_j 
 + \tilde{\nu}_{R i}^{\dagger} 
   \left( m^2_{ \tilde{\nu}} \right)_{ij}
    \tilde{\nu}_{R j} 
 + \tilde{e}_{R i}^{\dagger} 
   \left( m^2_{ \tilde{e}} \right)_{ij} 
\tilde{e}_{R j}    \nonumber  \\ 
&+& m_{H_u}^2 H_u^{\dagger} H_u + m_{H_d}^2 H_d^{\dagger} H_d  
+ \left(  B \mu H_d H_u 
 + \frac{1}{2} B_{\nu} M_{R_{ij}} 
 \tilde{\nu}_{R i}^{\dagger} \tilde{\nu}_{R j} 
+ h.c. \right)  \nonumber \\ 
&+& \left( 
  A_{\nu}^{ij} \tilde{\nu}_{R i}^{\dagger}  \tilde{\ell}_j H_u 
+ A_e^{ij} \tilde{e}_{R i}^{\dagger} \tilde{\ell}_j H_d  +h.c.  
 \right)  \nonumber \\ 
&+& \left( 
    \frac{1}{2} M_1 \tilde{B}  \tilde{B}  
 +  \frac{1}{2} M_2 \tilde{W}^a  \tilde{W}^a  
  + \frac{1}{2} M_3 \tilde{G}^a  \tilde{G}^a  + h.c. \right) .
 \label{softterms} 
\end{eqnarray}

Current experimental results severely constrain 
 soft SUSY breaking parameters to be almost flavor blind and real. 
In our analysis, we adopt the most commonly considered scenario, 
 the constraint MSSM (CMSSM), and impose the universal boundary conditions 
 for soft SUSY breaking parameters at the GUT scale: 
\begin{eqnarray}
& & \left( m^2_{\tilde{\ell}} \right)_{ij} 
= \left( m^2_{ \tilde{\nu}} \right)_{ij}
=   \left( m^2_{ \tilde{e}} \right)_{ij} = m_0^2 \delta_{ij} \; , 
 \nonumber \\ 
& &m_{H_u}^2=m_{H_d}^2 = m_0^2  \; , 
  \nonumber \\ 
& & A_{\nu}^{ij} = A_0 Y_{\nu}^{ij}\; , \; \; 
  A_{e}^{ij} = A_0 Y_{e}^{ij} \; , 
  \nonumber \\ 
& & M_1=M_2=M_3= M_{1/2} \; ,  
\end{eqnarray}
and extrapolate the soft SUSY breaking parameters 
 to the electroweak scale according to their 
 renormalization group equations (RGEs). 
Although the boundary conditions are flavor-blind, 
 the LFV terms in the soft SUSY breaking parameters 
 such as the off-diagonal components of 
 $\left( m^2_{\tilde{\ell}} \right)_{ij}$ and $A_{e}^{ij}$ 
 are induced by the LFV interactions, namely, 
 the neutrino Dirac Yukawa couplings. 
For example, the LFV effect most directly emerges 
 in the left-handed slepton mass matrix through the RGEs such as 
\begin{eqnarray}
\mu \frac{d}{d \mu} 
  \left( m^2_{\tilde{\ell}} \right)_{ij}
&=&  \mu \frac{d}{d \mu} 
  \left( m^2_{\tilde{\ell}} \right)_{ij} \Big|_{\mbox{MSSM}} 
 \nonumber \\
&+& \frac{1}{16 \pi^2} 
\left( m^2_{\tilde{\ell}} Y_{\nu}^{\dagger} Y_{\nu}
 + Y_{\nu}^{\dagger} Y_{\nu} m^2_{\tilde{\ell}} 
 + 2  Y_{\nu}^{\dagger} m^2_{\tilde{\nu}} Y_{\nu}
 + 2 m_{H_u}^2 Y_{\nu}^{\dagger} Y_{\nu} 
 + 2  A_{\nu}^{\dagger} A_{\nu} \right)_{ij}  \; ,
 \label{RGE} 
\end{eqnarray}
where the first term in the right hand side denotes 
 the normal MSSM term with no LFV. 
In the leading-logarithmic approximation, 
 the off-diagonal components ($i \neq j$)
 of the left-handed slepton mass matrix are estimated as 
\begin{eqnarray}
 \left(\Delta  m^2_{\tilde{\ell}} \right)_{ij}
 \sim - \frac{3 m_0^2 + A_0^2}{8 \pi^2} 
 \left( Y_{\nu}^{\dagger} L Y_{\nu} \right)_{ij} \; ,  
 \label{leading}
\end{eqnarray}
where the distinct thresholds of the right-handed 
 Majorana neutrinos are taken into account 
 by the matrix $ L = \log [M_{GUT}/M_{R_i}] \delta_{ij}$.

The effective Lagrangian relevant for the LFV processes 
 ($\ell_i \rightarrow \ell_j \gamma$) is described as 
\begin{eqnarray}
 {\cal L}_{\mbox{eff}}= 
 -  \frac{e}{2} m_{\ell_i} \overline{\ell}_j \sigma_{\mu \nu} F^{\mu \nu} 
 \left(A_L^{j i} P_L + A_R^{j i} P_R \right) \ell_i  \; , 
\end{eqnarray}
where $P_{R, L} = (1 \pm \gamma_5)/2 $ is  
 the chirality projection operator, 
 and  $A_{L,R}$ are the photon-penguin couplings of 1-loop diagrams 
 in which chargino-sneutrino and neutralino-charged slepton 
 are running. 
The explicit formulas of $A_{L,R}$ etc. used in our analysis 
 are summarized in \cite{Hisano-etal}. 
The rate of the LFV decay of charged-leptons is given by 
\begin{eqnarray}
\Gamma (\ell_i \rightarrow \ell_j \gamma) 
= \frac{e^2}{16 \pi} m_{\ell_i}^5 
 \left( |A_L^{j i}|^2  +  |A_R^{j i}|^2  \right).    
\end{eqnarray}
The following approximation formula \cite{Hisano-etal} 
 is useful to understand the order of magnitude 
 of the LFV decay rate:  
\begin{eqnarray}
\Gamma (\ell_i \rightarrow \ell_j \gamma) 
 \sim  \frac{e^2}{16 \pi} m_{\ell_i}^5 
 \times  \left(\frac{\alpha_2}{4 \pi} \right)^2  
 \frac{| \left(\Delta  m^2_{\tilde{\ell}} \right)_{ij}|^2}{M_S^8} 
 \tan^2 \beta \; , 
 \label{LFVrough}
\end{eqnarray}
where $M_S$ is the average slepton mass at the electroweak scale, 
 and $ \left(\Delta  m^2_{\tilde{\ell}} \right)_{ij}$ 
 is the slepton mass estimated in Eq.~(\ref{leading}). 
 From these formulas, once we obtain the information 
 of the neutrino Dirac Yukawa coupling matrix 
 and the right-handed neutrino masses, 
 we can predict the LFV decay rate, 
 for a given set of universal boundary conditions 
 and $\tan \beta$.

Recent various cosmological observations, in particular, 
 the Wilkinson Microwave Anisotropy Probe (WMAP) satellite \cite{WMAP},  
 have established the $\Lambda$CDM cosmological model 
 with a great accuracy. 
The relic abundance of the cold dark matter (CDM) in 2$\sigma$ range 
 has been measured as
\begin{eqnarray}
 \Omega_{CDM} h^2 = 0.1131 \pm 0.0034 . 
\label{WMAP} 
\end{eqnarray}
As is well-known, in SUSY models with the $R$-parity conservation, 
 a neutralino, if it is the lightest sparticle (LSP), 
 is a promising candidate for the CDM in the present universe. 
When we apply the WMAP result to the relic density of 
 the neutralino LSP, the parameter space of the CMSSM 
 is dramatically reduced into the narrow stripe   
 due to the great accuracy of the WMAP data  \cite{CDM}. 
We take into account this cosmological constraint 
 in our analysis of the LFV processes below. 
For $\tan \beta=45$, $\mu>0$ and $A_0=0$, for example, 
 we can find the approximate relation between $m_0$ and $M_{1/2}$ 
 such as 
\begin{eqnarray}
\left( \frac{m_0}{1 \; \mbox{GeV}} \right) 
=  125.3 
+ 0.329 \left(\frac{M_{1/2}}{1 \; \mbox{GeV}} \right) 
+ 5 \times 10^{-5} 
 \left( \frac{M_{1/2}}{1 \; \mbox{GeV}} \right)^2,  
\label{relation} 
\end{eqnarray} 
along which the observed relic abundance, 
 $\Omega_{CDM} h^2 = 0.113$, is realized. 
In our analysis, we have employed 
 the SOFTSUSY 3.1.4 package \cite{softsusy} 
 to solve the MSSM RGEs and produce mass spectrum. 
Then, the relic abundance of the neutralino LSP 
 is calculated by using the micrOMEGAs 2.4 \cite{micromega} 
 with the output of SOFTSUSY in the SLHA format \cite{slha}.

Now let us consider two example models in the class of 
 predictive SO(10) models. 
The first example is the minimal SUSY SO(10) model 
 analyzed in \cite{Fukuyama-Okada}, in which only one {\bf 10} and 
 one $\overline{\bf 126}$ Higgs multiplets have Yukawa couplings 
 with {\bf 16} matter multiplets such as 
\begin{eqnarray}
 W_Y = Y_{10}^{ij} {\bf 16}_i H_{10} {\bf 16}_j 
           +Y_{126}^{ij} {\bf 16}_i H_{126} {\bf 16}_j , 
\label{Yukawa1}
\end{eqnarray} 
where ${\bf 16}_i$ is the matter multiplet of the $i$-th generation,  
 $H_{10}$ and $H_{126}$ are the Higgs multiplet of {\bf 10} 
 and $\overline{\bf 126} $ representations under SO(10), respectively. 
By virtue of the gauge symmetry, the Yukawa couplings, 
 $Y_{10}$ and $Y_{126}$, are, in general, complex symmetric 
 $3 \times 3$ matrices.

We assume some appropriate Higgs multiplets, 
 whose vacuum expectation values (VEVs) correctly 
 break the SO(10) gauge symmetry into the Standard Model one 
 at the GUT scale, $M_{GUT} \simeq  2 \times 10^{16}$ GeV. 
Note that $H_{10}$ and $H_{126}$, respectively, includes 
 one pair of up-type and down-type Higgs doublets 
 in the same representation as the pair in the MSSM. 
Suppose that the Higgs doublets in the MSSM are realized 
 as one light linear combination of the two pairs of Higgs 
 doublets and the other combination is as heavy as the GUT scale,
 realizing the MSSM as the low energy effective theory. 
The Higgs doublet in the representation 
 $(\overline{\bf 10}, {\bf 1}, {\bf 3})$ 
 under the Pati-Salam subgroup, 
 $G_{422}=$SU(4)$_c \times$ SU(2)$_L \times$ SU(2)$_R$, 
 is embedded in the $\overline{\bf 126} $ representation, 
 whose nonzero VEV breaks the Pati-Salam subgroup 
 to the SM one and at the same time, generates 
 the masses of the right-handed neutrinos. 
Providing suitable VEVs for the Higgs multiplets, 
 the quark and lepton mass matrices are characterized 
 by only two basic mass matrices, $M_{10}$ and $M_{126}$,   
 and three complex coefficients, $c_{10}$, $c_{126}$ and $c_R$, 
 at the GUT scale:  
\begin{eqnarray}
  M_u &=& c_{10} M_{10} + c_{126} M_{126}   \nonumber \\
  M_d &=&     M_{10} +     M_{126}   \nonumber \\
  M_D &=& c_{10} M_{10} -3 c_{126} M_{126}   \nonumber \\
  M_e &=&     M_{10} -3     M_{126}   \nonumber \\
  M_R &=& c_R M_{126}  \; , 
 \label{massmatrix1}
\end{eqnarray} 
where $M_u$, $M_d$, $M_D$, $M_e$, and $M_R$ 
 denote the up-type quark, down-type quark, 
 neutrino Dirac, charged-lepton, and 
 right-handed neutrino Majorana mass matrices, respectively. 
Except for $c_R$, 
 which is used to determine the overall neutrino mass scale, 
 this system has fourteen free parameters in total \cite{Matsuda-etal}, 
 and the strong predictability to the fermion mass matrices.

In Ref.~\cite{Fukuyama-Okada}, thirteen electroweak data 
 of six quark masses, three mixing angles and one phase 
 in the Cabibbo-Kobayashi-Maskawa (CKM) matrix, 
 and three charged-lepton masses are extrapolated to the GUT scale 
 according to the RGEs with a given $\tan \beta$, 
 and are used as inputs at the GUT scale. 
Solving the GUT mass matrix relation among quarks and charged-leptons 
 obtained by Eq.~(\ref{massmatrix1}), 
 we can describe the neutrino Dirac mass matrix $M_D$ 
 and $M_{126}$ as functions of only one free parameter $\sigma $, 
 the phase of the combination $(3_{10}+c_{126})/(-c_{10}+c_{126})$. 
It has been shown \cite{Fukuyama-Okada} that 
 the appropriate value of $\sigma $ and $c_R$ 
 can reproduce the light neutrino mass matrix 
 consistent with the observed neutrino oscillation data (as of 2002).  
For $\tan \beta=45$ and $\sigma=3.198$ fixed, 
 the right-handed Majorana neutrino mass eigenvalues 
 are found to be (in GeV) \cite{Fukuyama-Okada}
\begin{eqnarray}
 M_{R_1}=1.64 \times 10^{11},  \; 
 M_{R_2}=2.50 \times 10^{12},  \;  
 M_{R_3}=8.22 \times 10^{12},  
\label{MR1}
\end{eqnarray}
 where $c_R$ is fixed so that 
 $\Delta m_\oplus^2 = 2 \times 10^{-3} \mbox{eV}^2$. 
In the basis where both of the charged-lepton 
 and right-handed Majorana neutrino mass matrices 
 are diagonal with real and positive eigenvalues, 
 the neutrino Dirac Yukawa coupling matrix at the GUT scale 
 is unambiguously determined and explicitly given by 
\begin{eqnarray}
 Y_{\nu} = 
\left( 
 \begin{array}{ccc}
-0.000135 - 0.00273 i & 0.00113  + 0.0136 i  & 0.0339   + 0.0580 i  \\ 
 0.00759  + 0.0119 i  & -0.0270   - 0.00419  i  & -0.272    - 0.175   i  \\ 
-0.0280   + 0.00397 i & 0.0635   - 0.0119 i  &  0.491  - 0.526 i 
 \end{array}   \right) \; .  
\label{Ynu1}
\end{eqnarray}     
Although we will use the data in Eqs.~(\ref{MR1}) and (\ref{Ynu1}) 
 for our numerical analysis below, the neutrino oscillation parameters 
 given by these equations are, unfortunately, more than 3$\sigma$ 
 away from the current neutrino oscillation data \cite{NuData}. 
Some extension of the minimal SO(10) model is necessary 
 to improve the data fitting for the neutrino 
 oscillation parameters \cite{datafit}. 
However, the resultant neutrino Dirac Yukawa couplings 
 and the right-handed neutrino mass spectrum are not so much 
 changed by this improvement, and we believe that 
 our result for the LFV process in this paper gives, 
 at least, the correct order of magnitude 
 for the prediction of a slightly extended minimal SUSY SO(10) model.

As another example of predictive SO(10) models, 
 we consider a simple SUSY SO(10) GUT in five dimensions (5D)  
 proposed in \cite{F-O}, which can ameliorate several 
 theoretical problems in the minimal (renormalizable) SUSY SO(10) model 
 (see \cite{F-O} for detailed discussions on the theoretical problems 
  of the minimal SUSY SO(10) model). 
This model is defined in five dimensions with 
 the fifth dimension compactified on the 
 $S^1/(Z_2 \times Z_2^\prime)$ orbifold \cite{orbifoldGUT}. 
The SO(10) gauge symmetry in five dimensions is broken by 
 the orbifold boundary conditions to the Pati-Salam (PS) symmetry 
 SU(4)$_c\times$ SU(2)$_L\times$SU(2)$_R$. 
All matter and Higgs multiplets are arranged to reside 
 only on a brane (PS brane) at one orbifold fixed point 
 where the PS gauge symmetry is manifest, 
 so that low energy effective four dimensional description 
 of this model is nothing but the PS model 
 with a special set of matter and Higgs multiplets. 
The PS symmetry is broken at the normal GUT scale, 
 $M_{GUT} \simeq 2 \times 10^{16}$ GeV, 
 in usual four dimensional manner, 
 while the full gauge coupling unification is 
 realized after incorporating threshold corrections 
 of Kaluza-Klein modes in the bulk gauge multiplets. 
Phenomenology of sparticles \cite{F-O2} and applications 
 of the model to cosmology such as inflation \cite{F-O3} 
 and baryogenesis via leptogenesis \cite{F-O4} 
 have been investigated.

In this model with a simple set of Higgs multiplets, 
 the structure of fermion Yukawa couplings are almost 
 the same as the one in the minimal SUSY SO(10) model. 
Assuming appropriate VEVs for Higgs multiplets, 
 fermion mass matrices are obtained as 
 (see \cite{F-O} for the detailed definition of the mass matrices 
  and the coefficients) 
\begin{eqnarray}
 M_u &=& c_{10} M_{1,2,2}+ c_{15} M_{15,2,2} \; , 
 \nonumber \\
 M_d &=& M_{1,2,2} + M_{15,2,2} \; ,   
 \nonumber \\
 M_D &=& c_{10} M_{1,2,2} - 3 c_{15} M_{15,2,2} \; , 
 \nonumber \\
 M_e &=& M_{1,2,2} - 3 M_{15.2,2} \; , 
 \nonumber \\
 M_R &=& c_R M_{10,1,3}, 
\label{massmatrix2}
\end{eqnarray} 
 which are characterized by three fundamental mass matrices 
 $M_{1,2,2}$, $M_{15,2,2}$ and $M_{10,1,3}$ 
 and complex coefficients $c_{10}$, $c_{15}$ and $c_R$. 
Comparing these expressions with Eq.~(\ref{massmatrix1}), 
 we can see that the combination of two mass matrices of 
 $M_{1,2,2}$ and $M_{15,2,2}$ among $M_u,M_d,M_D$, and $M_e$ 
 is the same as that of $M_{10}$ and $M_{126}$ 
 in the minimal SUSY SO(10) model and, therefore, 
 the procedure for fitting the realistic Dirac fermion 
 mass matrices is the same as in the minimal SO(10) model. 
On the other hand, $M_R$ is fully independent on the above 
 four Dirac fermion mass matrices, whereas 
 in the minimal SUSY SO(10) model it is described by 
 $M_{126}$ and not independent. 
This fact enables us to improve the data fitting 
 of the neutrino oscillation parameters.

Through the seesaw mechanism \cite{seesaw}, 
 the light neutrino mass matrix is given by
\bea
 m_\nu=Y_\nu^T M_R^{-1} Y_\nu v_u^2 = U_{MNS} D_\nu U_{MNS}^T 
\label{seesaw}
\eea
 in the basis where the mass matrix of charged lepton is diagonal 
 with real and positive eigenvalues.
Here $v_u$ is the VEV of the up-type Higgs doublet in the MSSM, 
 $D_\nu$ is the diagonal mass matrix of light neutrinos, 
 and $U_{MNS}$ is neutrino mixing matrix. 
This is equivalent to the expression of the right-handed neutrino 
 mass matrix as 
\bea 
 M_R = v_u^2 \left( 
  Y_\nu U_{TBM}^*D_\nu^{-1}U_{TBM}^\dagger Y_\nu^T  \right). 
\label{MRformula}
\eea 
Once the information of the Dirac Yukawa coupling, 
 the mass spectrum of the light neutrinos, 
 and the neutrino mixing matrix is obtained, 
 we can fix the right-handed neutrino mass matrix. 
In order to determine $M_R$, we follow the manner in \cite{F-O4}.

We consider the normal hierarchical case 
 for the light neutrino mass spectrum, for simplicity, 
 and describe $D_\nu$ in terms of the lightest mass eigenvalue 
 $m_1$ and the mass squared differences: 
\bea
 D_\nu=\mbox{diag}\left(m_1,~\sqrt{\Delta m_{12}^2 + m_1^2},
 ~\sqrt{\Delta m_{13}^2 + m_1^2}\right).  
\label{Dnu}
\eea 
Here we adopted the neutrino oscillation data \cite{NuData}: 
\bea 
  \Delta m_{12}^2=7.59\times 10^{-5}~\mbox{eV}^2,
~~\Delta m_{13}^2=2.43\times 10^{-3}~\mbox{eV}^2 
\label{nudata}
\eea
In addition, we assume the neutrino mixing matrix of 
 the so-called tri-bimaximal form \cite{TBM} 
\bea
 U_{TBM} = 
\left(
 \begin{array}{ccc}
 \sqrt{\frac{2}{3}}  & \sqrt{\frac{1}{3}} &  0 \\
-\sqrt{\frac{1}{6}}  & \sqrt{\frac{1}{3}} & \sqrt{\frac{1}{2}} \\ 
-\sqrt{\frac{1}{6}}  & \sqrt{\frac{1}{3}} & -\sqrt{\frac{1}{2}}
      \end{array} \right),
\label{TBM}
\eea
 which is in very good agreement with the current best fit 
 values of the neutrino oscillation data \cite{NuData}. 
As has been discussed above,  
 the data fit for the realistic charged fermion mass matrices  
 is the same as in the minimal SO(10) model, and we here use 
 the numerical value of $Y_\nu$ in Eq.~(\ref{Ynu1}) 
 at the GUT scale for $\tan \beta=45$, 
 in the basis where the charged lepton mass matrix is diagonal. 
Substituting Eqs.~(\ref{Ynu1}), (\ref{Dnu}), 
 (\ref{nudata}) and (\ref{TBM}) into Eq.~(\ref{MRformula}), 
 we obtain the right-handed neutrino mass 
 matrix as a function of only $m_1$. 
Since it has been shown \cite{F-O4} that the simple 5D SO(10) model 
 can reproduce the observed baryon asymmetry of the present universe 
 for $m_1 \simeq 1.8 \times 10^{-3}$ eV through non-thermal leptogenesis, 
 we take this value as a reference value for $m_1$. 
In this way, $M_R$ is now completely determined, but not 
 yet diagonalized. 
Changing the basis to diagonalize $M_R$, 
 we find the neutrino Dirac Yukawa matrix, 
\begin{eqnarray}
 Y_\nu = 
\left( 
 \begin{array}{ccc}
-0.00119 + 0.0000268i & -0.00108 - 0.000485 i & -0.000392 + 0.000421 i \\
 0.00135 + 0.00167 i  & -0.0253 + 0.00154 i & 0.0237 + 0.000851 i \\
-0.0265 - 0.0173 i    & 0.0609 + 0.0275 i & 0.790 - 0.0436i 
 \end{array}   \right) ,   
\label{Ynu2}
\end{eqnarray}   
 and the diagonal right-handed neutrino mass matrix 
 with eigenvalues (in GeV) 
\begin{eqnarray}
 M_{R_1}=1.03 \times 10^{10},  \; 
 M_{R_2}=7.55 \times 10^{11},  \;  
 M_{R_3}=3.22 \times 10^{15}.   
\label{MR2}
\end{eqnarray}

Now we are ready to analyze the $\mu \to e \gamma$ decay rate 
 by using the completely determined neutrino Dirac Yukawa 
 coupling matrix and the right-handed neutrino mass eigenvalues%
\footnote{
See \cite{FKO-LFV} for previous analysis 
 on the minimal SO(10) model with more general parameter sets. 
}, 
 Eqs.~(\ref{MR1}) and  (\ref{Ynu1}) for the minimal SO(10) model 
 while Eqs.~(\ref{Ynu2}) and  (\ref{MR2}) 
 for the simple 5D SO(10) model. 
For $\tan \beta =45$, $A_0=0$ and $\mu > 0$, 
 the branching ratio of $\mu \to e \gamma$ 
 for the minimal SO(10) model is shown in Fig.~1 
 as a function of the universal gaugino mass 
 at the GUT scale, along the relation of Eq.~(\ref{relation})
 to satisfy the observed relic density for the neutralino dark matter. 
The short dashed line corresponds to the branching ratio 
 of Eq.~(\ref{MEGsignal}) obtained from the MEG results 
 if the three events are considered as a positive signal, 
 while the long-dashed line to the upper limit of the branching 
 ratio set by the MEGA experiments. 
The universal gaugino mass $M_{1/2} \simeq 790$ GeV 
 (which corresponds to $m_0 \simeq 415$ GeV) 
 reproduces the MEG results.

Fig.~2 depicts the same result as in Fig.~1, but 
 for the simple 5D SO(10) model. 
In this case, the universal gaugino mass reproducing 
 $\BR (\mu \to e \gamma)=3 \times 10^{-12} $ is relatively light, 
 $M_{1/2} \simeq 420$ GeV (which corresponds to $m_0 \simeq 272$ GeV). 
This is because the components in the neutrino Dirac Yukawa matrix 
 relevant to the LFV between 1st and 2nd generations 
 are smaller than those in the minimal SO(10) model 
 and a lighter sparticle mass spectrum is necessary 
 in order to achieve the same branching ratio 
 (see Eq.~(\ref{LFVrough})).

For each model, we have pinned down the soft SUSY breaking mass
 parameters in the CMSSM, in the light of the MEG results. 
The set of the CMSSM parameters proposes a good benchmark point 
 for the SUSY search at the LHC, and we present 
 (sparticle) mass spectra for each model in Table~1, 
 along with other observables which can be compared 
 with the current experimental bounds. 
In both results, the lower bound on the Higgs boson mass 
 $m_h \geq 114.4$ GeV \cite{Higgs} is satisfied. 
Other phenomenological constraints we consider here are 
\begin{eqnarray}
& 2.85 \times 10^{-4} \leq \BR(b \rightarrow s  \gamma) \leq 4.24
\times 10^{-4} \; (2 \sigma ) & \hspace{1cm} \cite{bsgamma} 
\label{bsg}, \\
&  \BR(B_{s} \rightarrow \mu^{+} \mu^{-} ) < 5.8 \times 10^{-8}  &
\hspace{1cm} \cite{bsmumu} \label{bsmm}. 
\end{eqnarray}
The results for the minimal SO(10) model satisfy these constraints, 
 while $\BR(b \rightarrow s  \gamma)$ for the simple 5D SO(10) model 
 is marginal (about 3.4$\sigma$ away from the center value).

The muon anomalous magnetic dipole moment (muon aMDM) 
 has been measured in a great precision as \cite{E821} 
\begin{eqnarray}
  a_\mu^{\mbox{exp}} = 11659208.0(6.3)\times 10^{-10},
\label{amuexp}
\end{eqnarray}
 where the number in parentheses shows $1\sigma$ uncertainty. 
On the other hand, the SM predictions were calculated 
 \cite{Davier:2009ag} (see also
 \cite{de Troconiz:2004tr,Hagiwara:2006jt, 
       DEHZ,Jegerlehner:2007xe,Prades:2009qp}),  
\begin{eqnarray}
 a_\mu^{\mbox{SM}}[\tau] &=& 11659193.2(5.2)\times 10^{-10}, \nonumber \\
 a_\mu^{\mbox{SM}}[e^+e^-] &=& 11659177.7(5.1)\times 10^{-10},
\label{amuSM}
\end{eqnarray}
 by using data in the hadronic $\tau$ decay and 
 $e^+e^-$ annihilation to hadron, respectively, 
 in calculating the hadronic contributions to 
 the muon aMDM, $a_\mu^{\mbox{SM}}[\tau]$ and
 $a_\mu^{\mbox{SM}}[e^+e^-]$. 
The deviations of the SM predictions from the experimental result
 are given by 
\begin{eqnarray}
\Delta a_\mu[\tau]
&\equiv& a_\mu^{\mbox{exp}} - a_\mu^{\mbox{SM}}[\tau]
 = 14.8(8.2)\times 10^{-10}, \nonumber \\ 
\Delta a_\mu[e^+e^-]
 &\equiv& a_\mu^{\mbox{exp}} - a_\mu^{\mbox{SM}}[e^+e^-]
 = 30.3(8.1)\times 10^{-10}, 
\label{deviation}
\end{eqnarray}
 which correspond to $1.8\sigma$ and $3.7\sigma$ deviations, 
 respectively. 
These deviations may be from the sparticle contributions. 
Table~1 also include the sparticle contributions to $\Delta a_\mu$. 
We can see that the result for the minimal 5D SO(10) model favors 
 the deviation obtained by using the data of the hadronic
 $\tau$ decay, while the deviation from the $e^+ e^-$ data 
 is favored by the result for the simple 5D SO(10) model.

With $M_{1/2}$ (and $m_0$) determined in each SO(10) model, 
 we can also predict the LFV decay rates of tau lepton 
 in the same analysis as for $\mu \to e \gamma$. 
For the minimal SO(10) model, we find 
\bea 
 {\rm BR}(\tau \to \mu \gamma) \simeq 5.74 \times 10^{-10} 
\; \; {\rm and} \; \;
 {\rm BR}(\tau \to e \gamma)   \simeq 1.32 \times 10^{-10}, 
\eea 
 while for the simple 5D SO(10) model  
\bea 
 {\rm BR}(\tau \to \mu \gamma) \simeq 5.53 \times 10^{-10}
\; \; {\rm and} \; \;
 {\rm BR}(\tau \to e \gamma) \simeq 1.34 \times 10^{-10}.    
\eea  
These predicted values are well below 
 the current experimental upper bounds \cite{tauLFV}, 
${\rm BR}(\tau \to \mu \gamma) \leq 4.4 \times 10^{-8}$ 
 and 
${\rm BR}(\tau \to e \gamma) \leq 3.3 \times 10^{-8}$. 
Note that improvements by one or two orders of magnitude 
 are expected at future super B-factories \cite{superB} 
 and the branching ratio of order $10^{-10}$ can be 
 tested in the near future.

In summary, recent MEG results of a search 
 for the $\mu \to e \gamma$ decay might show a sign of 
 a certain new physics beyond the Standard Model. 
The primary candidate for new physics which can give 
 the rate of the LFV decay accessible to the MEG experiment 
 is supersymmetric models. 
A certain class of SUSY SO(10) GUT models (predictive SO(10) models) 
 can unambiguously determine fermion Yukawa coupling matrices, 
 in particular, the neutrino Dirac Yukawa matrix 
 and the mass spectrum of the right-handed neutrinos. 
As an example of such models, we have considered the minimal SO(10) 
 model and a simple 5D SO(10) model. 
The effective theory of the SO(10) models below the GUT scale 
 is described as the MSSM with the right-handed neutrinos. 
In these models, even if the universal boundary conditions 
 for the soft SUSY breaking parameters at the GUT scale, 
 LFV terms are induced in the sparticle sector through 
 the RGE effects with the neutrino Dirac Yukawa matrix, 
 which give rise to a sizable rate of the $\mu \to e \gamma$ decay. 
Since the neutrino Dirac Yukawa matrix and the mass spectrum 
 of the right-handed neutrinos are completely determined, 
 the LFV rate can be predicted once the set of input parameters 
 in the CMSSM is given. 
Requiring to reproduce $\BR(\mu \to e \gamma) = 3 \times 10^{-12}$ 
 given by the MEG results 
 and combining this requirement with the cosmological constraint
 on the observed relic abundance of neutralino dark matter, 
 we have pinned down the parameter set in the CMSSM and then, 
 have proposed benchmark points for each SO(10) model. 
We have checked other phenomenological constraints 
 for each benchmark point. 
Finally, our strategy in this paper is applicable 
 also to other SUSY (GUT) models if they can completely determine 
 the neutrino Dirac Yukawa matrix and the right-handed 
 neutrino mass spectrum. 
Depending on models, different benchmark points can be  
 pinned down, which will be tested at the LHC.

\section*{Acknowledgments}
We are grateful to Hieu Minh Tran for his help 
 in some of numerical analysis. 
This work is supported in part by 
 the Grant-in-Aid for Scientific Research from the Ministry 
 of Education, Science and Culture of Japan, 
 \#20540282 and \#21104004 (T.F.), 
 and by the DOE Grants, \# DE-FG02-10ER41714 (N.O.).


\newpage 

\begin{figure}
\begin{center}
\epsfig{file=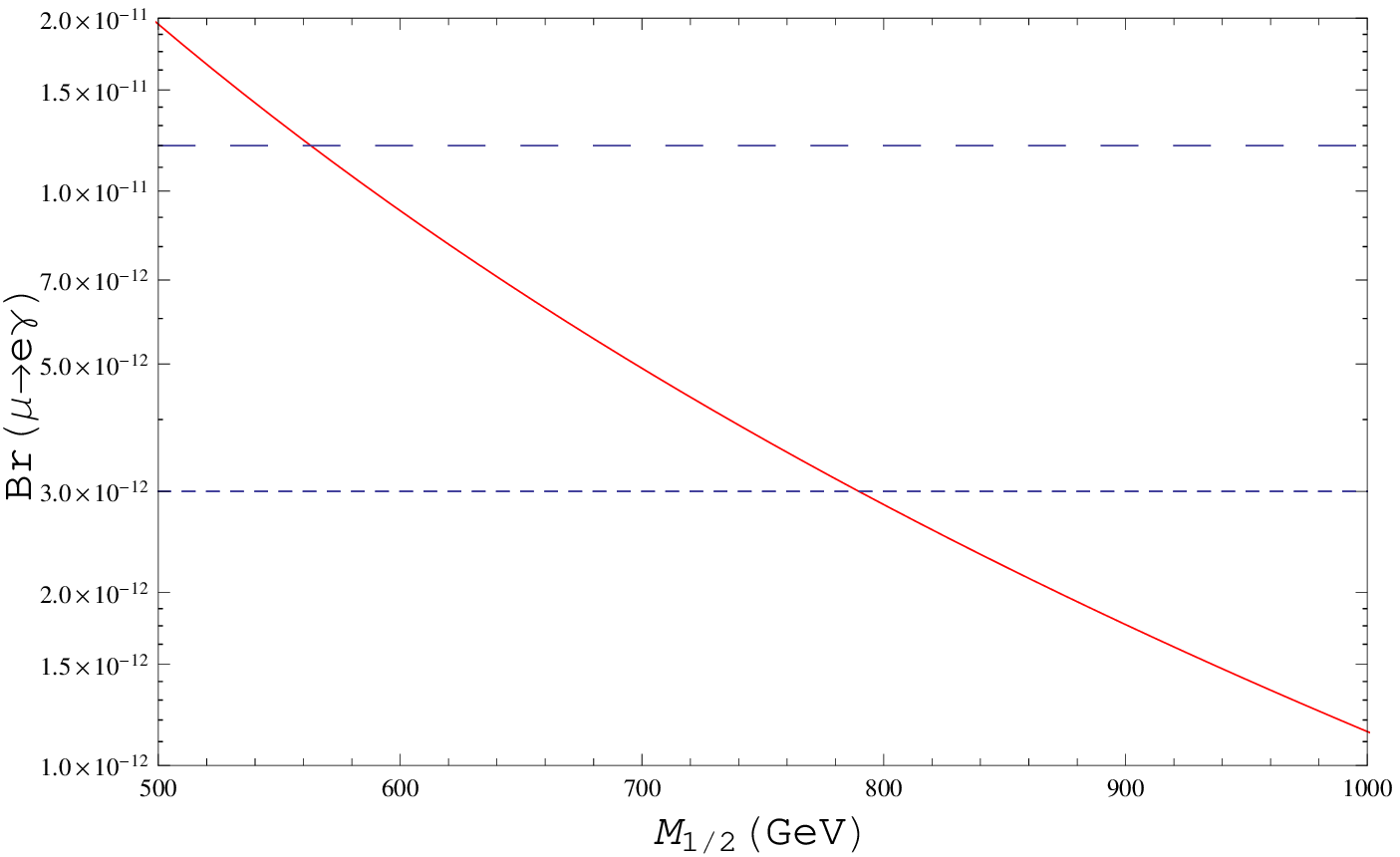, width=12cm}
\caption{ 
The branching ratio $\BR(\mu \to e \gamma)$ for the minimal SO(10) model 
 as a function of $M_{1/2}$ (GeV) 
 along the cosmological constraint of Eq.~(\ref{relation}). 
The short dashed line corresponds to the MEG result, 
 $\BR(\mu \to e \gamma)=3 \times 10^{-12}$ 
 when three evens are considered as a positive signal, 
 while the long dashed line to the upper limit on 
 $\BR(\mu \to e \gamma)=1.2 \times 10^{-11}$ 
 given by the MEGA experiment. 
}
\end{center}
\end{figure}
\begin{figure}
\begin{center}
\epsfig{file=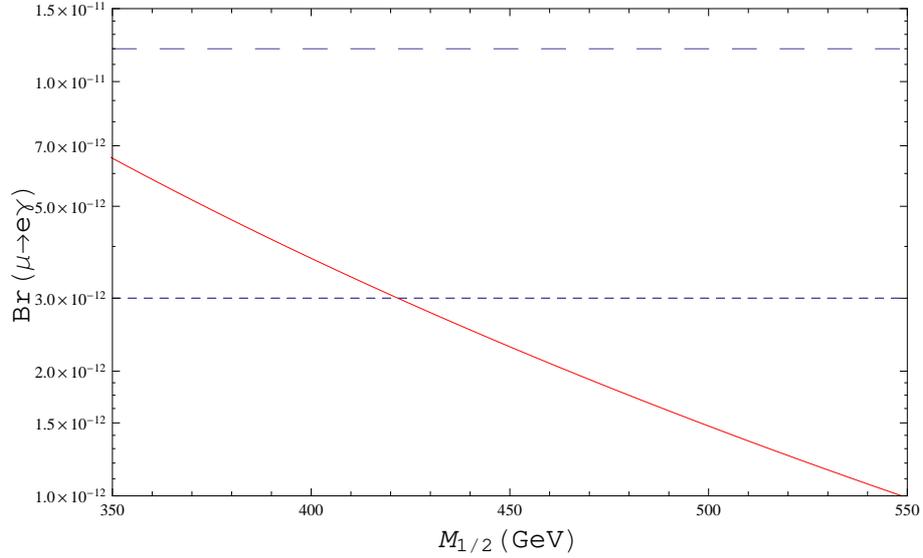, width=12cm}
\caption{ 
The same figure as Fig.~1 but for the simple 5D SO(10) model. 
}
\end{center}
\end{figure}

\newpage


\begin{table}
\caption{
Mass spectra (in GeV) and phenomenological constraints 
for the two SUSY SO(10) models 
with the universal boundary conditions in the CMSSM 
} \label{table}
\begin{center}
\begin{math}
\begin{array}{|c|c|c|}
\hline  & $minimal SO(10) model$ & $simple 5D SO(10) model$ \\
\hline m_0     & 415   & 272  \\ 
       M_{1/2} & 790   & 420  \\ 
       A_0     &   0   &   0  \\ 
    \tan \beta &  45   &   45 \\ 
\hline 
 h_0   & 119 & 115  \\ 
 H_0   & 786 & 449  \\ 
 A_0   & 787 & 449  \\
 H^\pm & 791 & 457  \\  
\hline 
\tilde{g} & 1756 & 981 \\ 
{\tilde{\chi}^0}_{1,2,3,4} & 
  333, 631, 928, 938 & 171, 324, 535, 548 \\ 
{\tilde{\chi}^{\pm}}_{1,2} & 
  631, 938 & 324, 548 \\
\hline 
\tilde{d},\tilde{s}_{R,L} & 1576, 1645 & 898, 934 \\
\tilde{u},\tilde{c}_{R,L} & 1582, 1643 & 901, 931 \\
\tilde{b}_{1,2} & 1409, 1473 & 784, 849 \\
\tilde{t}_{1,2} & 1266, 1475 & 698, 864 \\
\hline 
\tilde{\nu}_{e,\mu,\tau} 
 & 667, 667, 619 & 386, 386, 355 \\
\tilde{e},\tilde{\mu}_{R,L} 
 & 511, 672 & 317, 395  \\
\tilde{\tau}_{1,2} 
 & 342, 642 & 186, 392 \\
\hline 
\BR(b \rightarrow s \gamma) & 
 3.27 \times 10^{-4} & 2.36 \times 10^{-4} \\
\BR(B_s \rightarrow \mu^+ \mu^-) & 
 1.04 \times 10^{-8} & 4.95 \times 10^{-9} \\ 
\Delta a_{\mu} & 
 12.0 \times 10^{-10} & 37.7 \times 10^{-10} \\
\hline \Omega h^2 & \multicolumn{2}{|c|}{0.113}  \\
\hline 
\end{array}
\end{math}
\end{center}
\end{table}

\end{document}